\documentclass{eas}
\usepackage{graphicx}
\begin{document}

\TitreGlobal{Mass Profiles and Shapes of Cosmological Structures}

\title{Tracing the mass profiles of galaxy clusters with member galaxies}
\author{Andrea Biviano}\address{INAF -- Osservatorio Astronomico di Trieste, 
via G.B. Tiepolo 11, I-34131 Trieste, Italy -- biviano@ts.astro.it}
\runningtitle{Cluster mass profiles with member galaxies}
\setcounter{page}{23}
\index{Biviano, A.}

\begin{abstract}
The mass distribution of galaxy clusters can be determined from the
study of the projected phase-space distribution of cluster galaxies.
The main advantage of this method as compared to others, is that it
allows determination of cluster mass profiles out to very large
radii. Here I review recent analyses and results on this topic.  In
particular, I briefly describe the Jeans and Caustic methods, and the
problems one has to face in applying these methods to galaxy
systems. Then, I summarize the most recent and important results on
the mass distributions of galaxy groups, clusters, and superclusters.
Additional covered topics are the relative distributions of the dark
and baryonic components, and the orbits of galaxies in clusters.
\end{abstract}

\maketitle

\section{Introduction}
Knowledge of the mass distribution within clusters, (also in relation
to the distributions of the different cluster components), gives
important clues about the formation process of the clusters and of the
galaxies in them, as well as on the nature of dark matter. There have
been many studies of the mass distribution in galaxy systems over the
last decade, stimulated by the discovery of the 'universal' 'NFW' mass
density profile of dark matter haloes by Navarro et al. (1996, 1997).

A cluster mass distribution can be derived in three ways: 1) through
the gravitational lensing of distant objects, 2) using the spatial
distribution and temperature profile of the X-ray emitting,
intra-cluster (IC hereafter) gas, and 3) through the kinematics and
spatial distribution of 'tracer particles' moving in the cluster
potential. Lensing mostly works for clusters at intermediate and large
redshifts, and only few nearby cluster lenses are known (see
e.g. Cypriano et al. 2004). X-ray observations only sample the inner
(see, e.g., Pratt \& Arnaud 2002) or, at best, the virialized (Neumann
2005) cluster regions.  The third method is particularly suited for
studying the mass profiles of relatively nearby galaxy clusters, out
to large radii, well beyond the virialized region (see, e.g.,
Reisenegger et al. 2000). Additionally, using galaxy distributions it
is also possible to constrain the velocity anisotropies, and hence the
orbits, of cluster galaxies (see, e.g., Biviano \& Katgert 2004, BK04
hereafter).

In this paper, I review recent results on cluster mass profiles as
obtained from the analysis of the spatial and velocity distributions
of cluster member galaxies (see Biviano 2002 for another recent review
on this topic).  

\section{Methods \& Problems}
In order to determine a cluster mass profile, $M(<r)$, using the
projected phase-space distribution of its member galaxies we can use
the Jeans analysis (see, e.g., Binney \& Tremaine 1987, BT87
hereafter), or the 'Caustic' method recently introduced by Diaferio \&
Geller (1997; see also Diaferio 1999).

In the Jeans analysis the observable projected phase-space
distribution of galaxies is related to the cluster $M(<r)$, using the
Abel and Jeans equations (eqs. 4-55, 4-57, and 4-58 in BT97, for the
simplified case of a stable, non-rotating, spherically symmetric
system), through knowledge of the velocity anisotropy profile,
$\beta(r)$, which characterizes the orbits of cluster galaxies.

In the Caustic method, one infers $M(<r)$ of a given cluster from the
amplitude of the caustics in the plane of line-of-sight velocities
vs. projected clustercentric distances.  Formally, the caustic
amplitude is related to the gravitational potential through
a function ${\cal F}$ of the potential itself, and of $\beta(r)$, (see
eqs. 9 and 10 in Diaferio 1999). Numerical simulations indicate ${\cal
F} \approx \rm{const}$, but only at radii larger than the cluster
virial radius, $r_{200}$ (see Fig.~2 in Diaferio 1999). Hence, for
$r<r_{200}$ the Caustic mass estimate is not very accurate. On the
other hand, since the method does not rely on the assumption of
dynamical equilibrium, it is a very powerful tool to constrain $M(<r)$
at large radii.

Note that both with the Jeans and the Caustic method one samples
the {\em total,} not the {\em dark} mass of a cluster. This must be
taken into consideration when observational results are compared with
results from numerical simulations of collsionless dark matter haloes.
Using X-ray data it is possible to subtract the baryonic component of
the total mass and infer the dark mass distribution (see \L okas \&
Mamon 2003, LM03 hereafter), but this has not been done very often.

A fundamental problem of the Jeans analysis (and, to a lesser extent,
also of the Caustic method) is the 'mass--orbit' degeneracy, i.e.  the
solution obtained for $\mbox{$M(<r)$}$ is degenerate with respect to the
solution obtained for $\beta(r)$. In order to break the degeneracy,
$\beta(r)$ must be constrained independently from $M(<r)$. This can be
achieved via the analysis of the {\em shape} of the galaxy velocity
distribution, that contains the required information about the orbital
anisotropy of cluster galaxies (Merritt 1987; van der Marel et
al. 2000, vdM00 hereafter; LM03). Alternatively, if several tracers of
the gravitational potential are available, the Jeans equation can be
solved for $M(<r)$ independently for each of the tracers, thus
restricting the range of possible solutions (BK04).

Another relevant problem in the Jeans analysis occurs if the cluster
is not in steady state and dynamical equilibrium.  Since clusters grow
by accretion of field galaxies (e.g. Moss \& Dickens 1977), they are
not steady-state systems.  Formally, inclusion of the time derivative
in the Jeans equation (BT87, eq. 4-29c) is then needed. On the other
hand, the fractional mass infall rate is estimated to be quite
negligible for nearby clusters (Ellingson et al. 2001). Moreover, most
of the mass is accreted in big, discrete clumps (Zabludoff \& Franx
1993). Hence, those clusters undergoing substantial mass accretion can
be identified through the presence of substructures in their galaxies
distribution, and excluded from the analysis (vdM00; Biviano \&
Girardi 2003). The problem gets tougher when one is dealing with small
galaxy systems (galaxy groups) most of which are probably still in a
pre-virialized collapse phase (Giuricin et al. 1988).

Interlopers are another potential problem in cluster mass estimates.
Cluster velocity dispersion estimates have been shown to be robust
vs. modifications of the method of interlopers removal (Girardi et
al. 1993), but estimates of the kurtosis of the velocity distribution
(needed to constrain $\beta$, see LM03) are not. Hence, it is
advisable to use other, more robust, estimators of the shape of a
cluster velocity distribution (e.g. the Gauss-Hermite moments, vdM00).
As the number of available redshifts for cluster galaxies increases,
the interloper selection procedure becomes more robust. Since in
practice spectroscopic samples of more than 100 member galaxies per
cluster are rare, quite often a 'composite' cluster is built by
stacking together the data for several clusters (see, e.g., Carlberg
et al. 1997; vdM00; Katgert et al. 2004). The procedure is justified
if clusters form a homologous family, as suggested by the existence of
a fundamental plane of cluster properties (Schaeffer et al. 1993;
Adami et al. 1998).

In the Jeans analysis it is generally assumed that clusters are
spherical and do not rotate. The composite cluster is spherically
symmetric by construction, and deviation from spherical symmetry is
unlikely to be a major problem for individual clusters either (see
vdM00; Sanchis et al. 2004). While evidence for cluster rotation has
been claimed for a couple of clusters (see, e.g., Biviano et al. 1996;
Dupke \& Bregman 2001), the energy content in the rotational component
is marginal. Finally, dynamical friction and galaxy mergers could in
principle invalidate the use of the {\em collisionless} Jeans
equation. However, the cluster velocity dispersion is too high for
galaxy mergers to take place, and the dynamical friction timescale is
too long for most cluster galaxies, except for very massive galaxies
(e.g. Biviano et al. 1992). These can be removed from the sample
(Katgert et al. 2004) prior to application of the Jeans analysis.
Galay mergers in low-velocity dispersion systems, like groups, can however
be a critical issue.

\section{Results}
\subsection{Small groups}
Since groups contain only a small number of galaxies each, their mass
profile can only be inferred by stacking several of them
together. Results so far are contradictory. Mahdavi et al. (1999) used
588 galaxies in 20 groups to conclude that their mass density profile
is consistent with a Hernquist (1990) model, and that the mass-to-galaxy
number-density profiles ratio is constant.  Mahdavi \& Geller (2004)
used the RASSCALS sample (893 galaxies in 41 nearby groups) to
determine a single power-law mass density profile, $\rho(r) \propto
r^{1.9 \pm 0.3}$ over a wide radial range (0--$2 \, r_{200}$).
Finally, Carlberg et al. (2001) analysed $\sim 800$ galaxies in $\sim
200$ groups from the CNOC2 survey, at redshifts $z=0.1$--0.55. They
found a cored mass density profile near the groups centres, decreasing
outside with a shallow slope, hence implying a steeply increasing
mass-to-light ratio with radius.

Despite these different results, all these studies conclude that
galaxies in groups have, on average, nearly isotropic orbits, except
late-type galaxies that are on mildly radial orbits (Mahdavi et
al. 1999).

\subsection{Clusters}
The Coma cluster, with its nearly 900 spectroscopically confirmed
cluster members (Adami et al. 2005), is probably the best studied
cluster in the Universe.  Leaving apart historical studies (see
Biviano 1998 and references therein), the first modern analysis of the
Coma cluster mass profile was done by Merritt \& Saha (1993). They
found that the density profile is cuspy near the centre, $\rho(r)
\propto r^{-2}$, and decreases as $\rho(r) \propto r^{-4}$, at large
radii, if the orbits are isotropic. More recently, LM03 found that the
NFW profile provides a good fit to the derived mass distribution, but
other models are acceptable, and even a cored profile near the centre
cannot be excluded. At variance with Merritt \& Saha, and unlike most
other studies, LM03 analysed the {\em dark,} not the total, mass
profile of the cluster. Using the kurtosis profile of the galaxies
velocity distribution, they obtained $-1.2 \leq \beta \leq 0.3$ for
early-type galaxies, i.e nearly isotropic orbits.

The above results are in agreement with those obtained through the
Caustic method (Geller et al. 1999; Rines et al. 2001). These studies
also constrained the behaviour of the mass density profile at very
large radii, $\rho(r) \propto r^{-\zeta}$ with $\zeta=3$--4. The
isothermal sphere model ($\zeta=2$) is rejected, while both a NFW and
a Hernquist (1990) model are acceptable. The mass-to-light ratio in
the $K$-band is found to be nearly constant out to $\simeq 6 \,
r_{200}$.

In a series of papers, Rines et al. (2000, 2003, 2004) have applied
the Caustic technique to another 8 nearby clusters of the CAIRNS
survey.  The results are not very different from those found for the
Coma cluster. Typically, $\rho(r) \propto r^{-1}$ near the centre, and
the best-fit asymptotic slope of the mass density profile at large $r$
is either $-3$ (NFW) or $-4$ (Hernquist 1990). When the NFW profile is
adopted, the best-fit values of the concentration parameter $c$ vary
between 5 and 17. Using the mass profiles obtained from the Caustic
technique in the Jeans analysis, $\beta \approx 0$ is obtained. The
mass-to-light ratio in the $K$ band is approximately constant out to
$r_{200}$, then decreases by a factor $\times 2$ out to the turnaround
radius.

Following up a previous study by Carlberg et al. (1997), vdM00 stacked
together 16 clusters at $z=0.17$--0.55 from the CNOC survey, to build a
composite cluster with $\simeq 990$ galaxies.  Through the analysis of
the shape of the galaxies velocity distribution, vdM00 were able to
constrain $-0.6 \leq \beta \leq 0.1$. From the Jeans analysis they
then determined a mass profile which is fully consistent with a
$c=4.2$ NFW model. The ratio between the mass density profile and the
galaxies number density profile was found to be nearly constant.

Biviano \& Girardi (2003) analysed a composite cluster of 1345 member
galaxies, obtained from the stacking of 43 clusters from the
2dFGRS. They performed a joint Jeans and Caustic analysis, using the
former method to determine the mass profile in the virialized region
($\leq r_{200}$), and the latter to extend $M(<r)$ to $2 \,
r_{200}$. I.e. they used each of the two methods in the cluster region
where it is expected to perform best. They found that a $c=5.6$ NFW
profile fits the data over the whole radial range explored.  Cored
profiles are also acceptable but only for core radii $< 0.1 \,
r_{200}$. The ratio of the mass density to the galaxy number density
profiles was found to be constant out to $r_{200}$, and to decrease
beyond that, mostly because of the contribution of late-type galaxies
in the external regions.

\begin{figure}[ht]
\centering
\includegraphics[width=12cm]{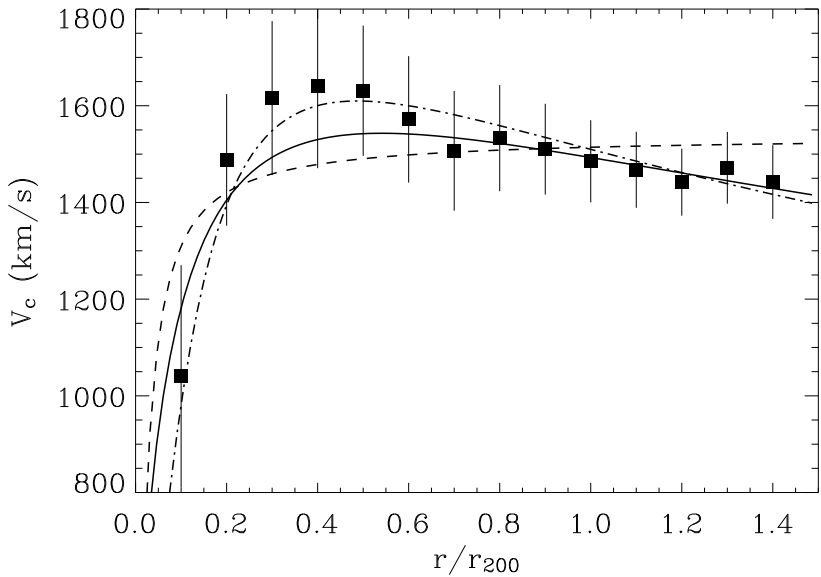}
\caption{The average circular velocity profile, 
$V_c \equiv \sqrt{G \, M(<r) \, r^{-1}}$,
of 59 clusters from the ENACS
(points with 1-$\sigma$ error bars).  The solid, dot-dashed, and dashed
lines are the best-fit NFW, Burkert (1995), and softened
isothermal sphere models, respectively}
\end{figure}

Katgert et al. (2004) analysed a composite of 1129 galaxies from 59
nearby clusters from the ENACS. $M(<r)$ was determined in a fully
non-parametric way.  Only early-type galaxies were used in the
analysis. Their velocity distribution was showed to imply nearly
isotropic orbits, hence $\beta \equiv 0$ was adopted in the Jeans
analysis. The resulting mass density profile has a slope of $-2.4 \pm
0.4$ at $r=r_{200}$, fully consistent with an asymptotic slope of $-3$
at larger radii. Both a $c=4 \pm 2$ NFW profile, and a Burkert (1995)
profile with a rather small core ($\leq 0.1 \, r_{200}$) are
acceptable (see Figure~1).  The ratio of the mass-to-luminosity
density profiles (in the $R$ band) is nearly constant out to $\sim 1.5
\, r_{200}$, when both the cD-like galaxies (that give an excess
luminosity near the centre), and the late-type galaxies (that give an
excess luminosity in the outer regions) are excluded.

The $M(<r)$ solution obtained for the early-type galaxies was later
confirmed by using a different tracer of the gravitational potential,
early-type {\em spirals} (Sa--Sb; BK04). Inverting the Jeans equation
(via the method of Solanes \& Salvador-Sol\'e 1990, developed from
Binney \& Mamon 1982), BK04 found that also early-type spirals have
nearly isotropic orbits in the cluster potential. On the other hand
the late spirals (Sbc--Sd) plus irregulars have $\beta \approx 0$ only
near the centre, and increasingly radial orbits outside. Galaxies in
substructures were found to have tangential orbits at all radii.

Work is in progress (Biviano \& Salucci in preparation) to determine the
relative distributions of the different mass components, namely the
baryonic matter (in galaxies and in the intra-cluster gas), the dark
matter in subhaloes, and the diffuse dark matter. Preliminary results
indicate that once the baryons and the dark matter subhaloes are
subtracted from the total mass profile, the remaining diffuse dark
matter profile is still consistent with both a Burkert (1995) and a
NFW model, but the concentration is a factor of two higher than
for the {\em total} mass profile. The galaxy baryonic mass profile is
found to be similar to the total mass profile, but the total baryonic
mass is more widely distributed, because of the dominant contribution
by the IC gas.

\subsection{Superclusters}
Superclusters are not virialized systems. It is then impossible to
determine their mass profile other than by the use of the Caustic
method. Reisenegger et al. (2000) and Rines et al. (2002) applied this
method to the Shapley and the A2197/A2199 supercluster, respectively.
The superclusters mass profiles are well fit by
both a NFW and a Hernquist (1990) model.

\section{Summary and perspectives}
Useful constraints can be put on the mass distribution within clusters
from the analysis of the projected phase-space distribution of their
member galaxies. The constraints are poor near the centre, $\rho(r)
\propto r^{\xi}$ with $-2 \leq \xi \leq 0$, because of an intrinsic
finite resolution (the size of the cD galaxy). Cored profiles are
allowed, but only for core sizes below this resolution size, $\leq 0.1
\, r_{200}$. Stronger constraints are obtained at large radii, and
they are stronger than those obtained with any other methods, $\rho(r)
\propto r^{\zeta}$ with $-4 \leq \zeta \leq -3$. In summary, the
isothermal sphere is ruled out, while the NFW, Burkert (1995) and
Hernquist (1990) models are acceptable.

When comparing a cluster mass density profile with the galaxy number
(or luminosity) density profiles, it must be realized that the result
of this comparison depends upon which cluster galaxy population (or,
almost equivalently, which photometric band) is considered. More
relevant is the comparison of the total mass distribution with the
baryonic mass distribution. The galaxy baryonic mass profile is similar
to the total mass profile, but the total baryonic mass distribution
(including the IC gas) is more extended.

As fas as the orbits are concerned, ellipticals, lenticulars, and
early-type spirals (Sa--Sb) are all found to be on nearly isotropic
orbits. Late-type spirals (Sbc-Sd) and irregulars have increasingly
radial velocity anisotropy with increasing radius, and galaxies in
substructures seem to be characterized by tangential orbits. The
implications of these results for the evolution of galaxies in
clusters is discussed in BK04.

Very little is known about the evolution of cluster dynamics with
redshift. The average mass profile and the orbital characteristics of
galaxies of nearby clusters (ENACS, CAIRNS, 2dFGRS) are similar to
those of medium-distant clusters (CNOC). Hence, little or no evolution
over $z \sim 0$--0.5 is implied (see also Girardi \& Mezzetti 2001).

In the near future it will be possible to use the SDSS products (see
e.g. Goto 2005) to reduce the current uncertainties on the dynamics of
nearby clusters.  It is also urgent to improve our understanding of
the dynamics of small galaxy systems (groups). A promising data-set in
this respect is the GEMS groups sample of Osmond \& Ponman (2004), for
which also X-ray temperatures are available. These can provide robust
scaling parameters for the build-up of a composite group sample.
Determining the average mass profile (and galaxy orbits) of distant
clusters ($z > 0.5$) is a more demanding task. Distant clusters are
likely to be in a more turbulent phase of their evolution than nearby
clusters, making the dynamical analysis even more difficult. As a
consequence, a deeper sampling of the projected phase-space
distribution of member galaxies is needed, with a strong investment in
terms of observing time (e.g. Demarco et al. 2005).

\vspace{0.5cm}

I wish to thank the organizers of the 2005 IAP meeting for inviting me
to give this review. I dedicate this paper to Patrizia, dolce amore.

\end{document}